\documentclass{article}
\usepackage{graphicx}
\usepackage{amsmath}

\input{tcilatex}

\begin{document}

\title{Optical Cavity Implementations of the Quantum Walk }
\author{Peter L. Knight\thanks{%
E-mail: p.knight@imperial.ac.uk}, Eugenio Rold\'{a}n\thanks{%
Permanent address: Departament d'\`{O}ptica, Universitat de Val\`{e}ncia,
Dr. Moliner 50, 46100--Burjassot, Spain. E-mail: eugenio.roldan@uv.es}, and
J. E. Sipe\thanks{%
Permanent address:\ Department of Physics, University of Toronto, Toronto
M5S 1A7, Canada. E-mail: sipe@physics.utoronto.ca} \\
%EndAName
Optics Section, Blackett Laboratory, Imperial College London, \\
London SW7 2AZ, United Kingdom}
\maketitle

\begin{abstract}
We discuss how the coined quantum walk on the line or on the circle can be
implemented using optical waves. We propose several optical cavity
configurations for these implementations.

\end{abstract}

\section{Introduction}

Classical implementations of quantum computing algorithms are interesting
for both practical and fundamental reasons. They allow actual
implementations that may serve as model systems for their quantum
counterparts, and they may shed some light on the profound differences
between classical and quantum waves.

The concept behind these classical implementations is simple: waves share
with quantum systems the possibility of creating superposition states \cite
{Spreeuw}. Of course, the entanglement of separated systems that is specific
to quantum mechanics is not present, but such entanglement does not seem to
play a central role in at least some of the proposed quantum computing
algorithms. This subject has been considered by a number of authors \cite
{Spreeuw}-\cite{Bhattacharya}. In particular, implementations of Grover's
quantum search algorithm have been proposed \cite{Kwiat}, and actually
implemented \cite{Bhattacharya}, using only classical resources. It must be
pointed out however that the lack of entanglement in the classical case
manifests through the exponential increase of the ammount of classical
resources needed with the increase of the problem size [1], and it would be
more appropriate to speak of simulations rather than actual implementations,
we shall come back to this point later.

The so--called quantum random walk (QW in the following) is a process that
has received a lot of attention in the quantum computing community \cite
{Aharonov}-\cite{Tregenna}; see the recent review by Julia Kempe \cite
{KempeCP}. The QW was introduced by Aharonov \textit{et al.} \cite{Aharonov}
ten years ago as the quantum analog of the classical random walk (RW for
short). The attention devoted to QWs is mainly motivated by the fact that as
some problems can be solved efficiently in classical computation with
algorithms based on RWs, there is a reasonable hope that QWs could give rise
to algorithms faster than classical for some specific problems. In work
along these lines, Kempe \cite{Kempe} showed that the hitting time of the
discrete QW from one corner of an $n$-bit hypercube to the opposite corner
is polynomial in the number of steps, $n$, whilst it is exponential in $n$
in the classical case. Subsequently Shenvi \textit{et al.} \cite{Shenvi}
showed that a QW can perform the same tasks as Grover's search algorithm,
and Childs \textit{et al.} \cite{Childs} have introduced an algorithm for
crossing a special type of graph exponentially faster than can be done
clasically. Moreover, as classical RWs are used in proofs of complexity to
place bounds on the difficulty of certain problems, then one might hope for
similar proofs for quantum algorithms using QWs \cite{Bartlett}.

QWs can be performed by a quantum computer, and a number of possible
implementations have been proposed: Travaglione \textit{et al.} \cite
{Travaglione} proposed its implementation in an ion trap quantum computer,
D\"{u}r \textit{et al.} \cite{Dür} consider the QW in an optical lattice,
and Sanders \textit{et al.} \cite{Sanders} have suggested its experimental
realization in cavity quantum electrodynamics. All these proposals consider
quantum systems. A different approach was followed by Zhao \textit{et al.} 
\cite{Zhao}, and subsequently by Jeong \textit{et al.} \cite{Jeong}, who
discussed the implementation of QWs using single photons and linear optical
elements. But it turns out that the quantum nature of light does not play
any role in the developement of the walk. That is, QWs can be performed
within classical optics. This has been noticed by Hillery \textit{et al.} 
\cite{Hillery} who have developed an interferometric analogy of QWs and made
a suggestion similar to that of Refs. \cite{Zhao,Jeong}. The problem with
these last proposals is that the number of optical elements grows quickly
with the number of steps in the QW.

In a recent paper \cite{Knight} we have shown that QWs can be understood as
an interference phenomenon (and thus as a \textit{classical} phenomenon in
the sense that quantization does not play any role in its understanding) and
can be clasically implemented in optical cavities. The advantage of the
optical cavity approach is that the resources needed (in the sense of the
number of optical elements) do not depend on the size of the implemented
walk. Moreover, we argued that the coined QW on the line had indeed been
optically implemented by Bouwmeester \textit{et al.}\cite{Bouwmeester} in
the context of an optical implementation of the Galton board, although the
authors did not explicitly make this link.

In the present paper we extend the results advanced in \cite{Knight} by
considering new cavity configurations, by proposing schemes for the
implementation of the QW on the circle, and by discussing in detail up to
what extent the experiment of Bouwmeester \textit{et al.} \cite{Bouwmeester}
constitutes an actual implementaion of the QW.

\section{The coined quantum walk}

In order to make this paper as self--contained as possible we briefly review
here some of the main characteristics of QWs.

Let us first consider the classical RW on the line. In a common version the
``walker'' -- the particle or system performing the RW -- takes one step to
the right or to the left depending on the (random) result of tossing a coin.
After $n$ steps, the probability of finding the walker at a distance $m$
from the origin is given by the binomial distribution which, for large $n$,
is well approximated by a Gaussian (dashed line in Fig. 1) whose standard
deviation is given by $\sqrt{n}$.

In the coined QW, the role of the coin is played by a qubit, that is, by a
two--level system. In analogy with its classical counterpart, the quantum
walker moves to the right or to the left depending on the internal state of
the qubit. The obvious difference with the classical RW is that the
coin--qubit can be set in a superposition state, and thus the walker can be
described by probability amplitudes associated with movement both to the
right and to the left. After each displacement of the walker, the state of
the qubit is set to a superposition state by means of a suitable unitary
transformation, typically (but not necessarily) the Hadamard transformation.
This operation plays, in a sense, the role of the toss of the coin in the
RW. Then a new displacement occurs, and so on and so forth.

\FRAME{fhFU}{5.8474cm}{4.2241cm}{0pt}{\Qcb{{\protect\small Probability
distribution for }$n=200${\protect\small \ for both the classical random
walk (dashed) and quantum walk (continuous). The initial conditions chosen
for calculating the QW were }$R_{0,0}=1/\protect\sqrt{2}${\protect\small \
and }$L_{0,0}=i/\protect\sqrt{2}${{\protect\small , see Eqs.(\ref{amp1},\ref
{amp2}). Notice that the quantum }}$P_{m}${\protect\small \ is null for odd }%
$m${\protect\small \ at odd }$n${\protect\small . We have represented only
nonzero values and joined them to guiding the eye. }}}{}{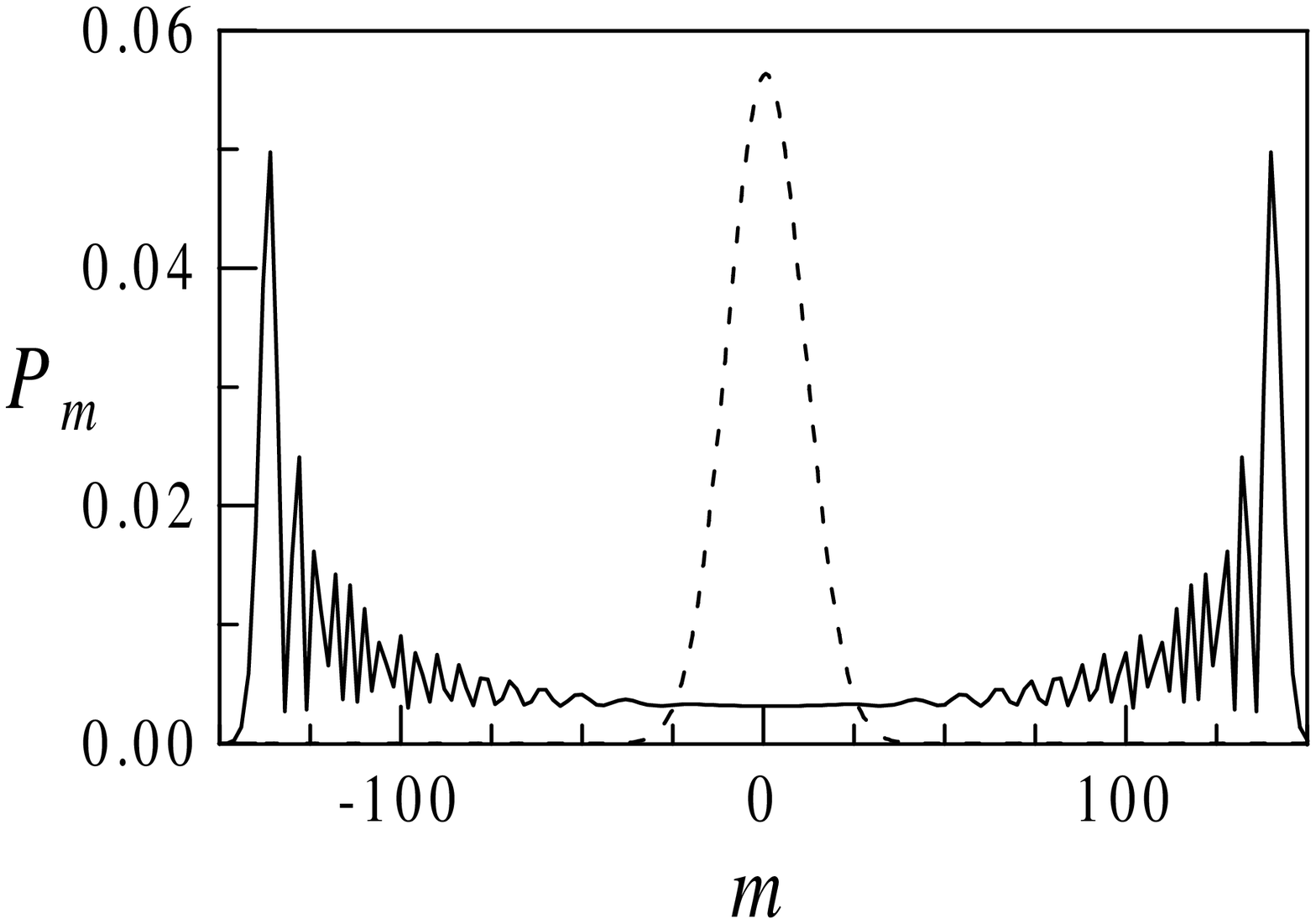}{\special{
language "Scientific Word"; type "GRAPHIC"; maintain-aspect-ratio TRUE;
display "USEDEF"; valid_file "F"; width 5.8474cm; height 4.2241cm; depth
0pt; original-width 7.76in; original-height 11.0653in; cropleft "0"; croptop
"1"; cropright "1"; cropbottom "0.4956"; filename 'fig1.ps';file-properties
"XNPEU";}}

The probability distribution in the quantum case is very different from the
classical one in several respects. It resembles the Airy function, as can be
seen in Fig.1 and we discuss elsewhere the asymptotics of the walk
distribution in terms of Airy functions \cite{Knight}. The standard
deviation of the QW probability distribution is linear in $n;$ that is, the
quantum walker walks quadratically faster than the classical walker. With
increasing time steps, the probability distribution is null for
alternatively even and odd values of $m$. Except for these nulls, the
probability distribution is quite flat and uniform in its central region.
Finally, we point out that the form of the probability distribution depends
strongly on the initial state of the coin. Several analytical approaches
have been developed that provide explicit solutions of the QW on the line 
\cite{Meyer,Nayak,Carteret}.

The QW on the circle is similar to the QW on the line. But the circle
contains only a finite number of discrete points from $m=0$ to $m=M-1$ and
then position $m=M+1$ is the same as position $0$ and so on. The probability
distribution is, of course, very different to that of the QW on the line (it
does not converge to a limit), and in Fig.2 we have plotted three examples
that are detailed in the caption (see also, \textit{e.g.,} \cite
{DAharonov,Bednarska}).

\FRAME{fhFU}{6.7568cm}{9.6212cm}{0pt}{\Qcb{{\protect\small Probability
distribution for the indicated number of steps, }$n$, {\protect\small for
the quantum walk on a circle with 61 discrete positions. The initial is }$%
R_{30,0}=1/\protect\sqrt{2}${\protect\small \ and }$L_{30,0}=i/\protect\sqrt{%
2}${\protect\small .}}}{}{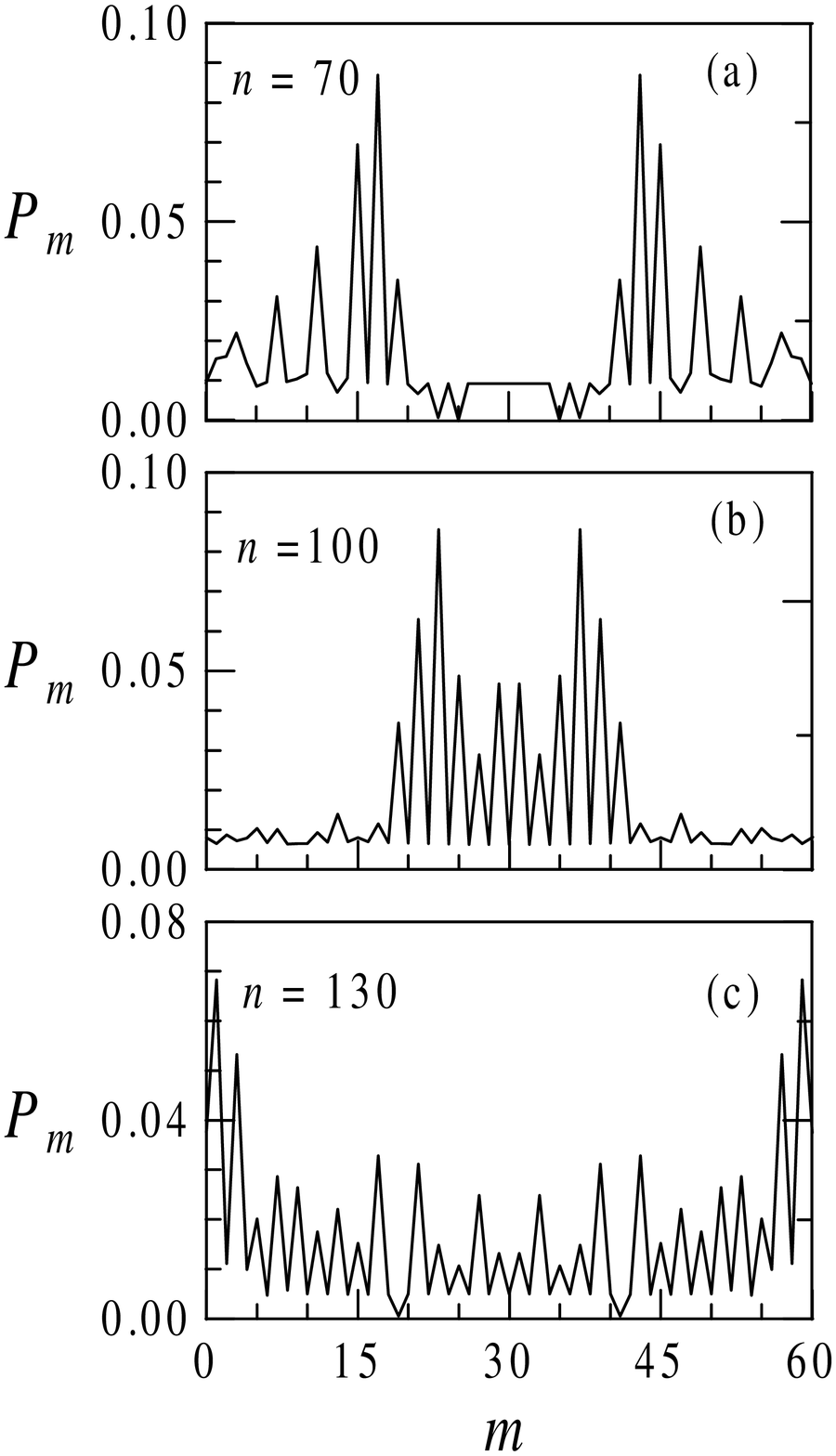}{\special{ language "Scientific Word";
type "GRAPHIC"; maintain-aspect-ratio TRUE; display "USEDEF"; valid_file
"F"; width 6.7568cm; height 9.6212cm; depth 0pt; original-width 19.7103cm;
original-height 28.1058cm; cropleft "0"; croptop "1"; cropright "0.9987";
cropbottom "0"; filename 'fig2.ps';file-properties "XNPEU";}}So-called
quantum random walks are of course not random, since the evolution of the
quantum walk is completely deterministic, and randomness enters only when a
position measurement of the walker is finally performed. Thus we refer to
this process simply as a ``quantum walk.'' We mention that a different but
related type of quantum walk, the so--called continuous QW, was introduced
by Farhi and Gutmann \cite{Farhi}. The continuous QW does not need the use
of any coin or qubit. We refer the interested reader to the literature \cite
{Farhi,KempeCP} for the details of this walk, which we will not consider
here.

\section{Classical implementation of the quantum walk in optical cavities}

\subsection{The Hadamard walk}

Let us first consider the implementation of the standard Hadamard walk. In
our approach, the role of the walker will be played by the frequency of a
light field, and the role of the coin will be played by some other degree of
freedom of the light field. This can be its spatial path or its polarization
state. That is, in the terminology of \cite{Spreeuw}, we consider both
position and polarization \textit{cebits.} A cebit is a two--component
complex vector representing the light field, and is the classical
counterpart of the qubit.

For example, the light field 
\begin{equation}
\vec{E}=\sum_{m=-l}^{l}\vec{E}_{m}e^{i\left[ \left( \omega_{0}+m\bar{\omega }%
\right) t-k_{m}z\right] }+c.c.,
\end{equation}
where $\omega_{0}$ is the carrier frequency, $\bar{\omega}$ is the frequency
difference between successive frequency components, and $z$ is the
propagation direction (the vector $\vec{E}_{m}$ lies in the $\left(
x,y\right) $ plane), can be represented by the abstract state 
\begin{equation}
\left| \psi\right) =\sum_{m=-l}^{l}\left[ R_{m}\left| m,x\right)
+L_{m}\left| m,y\right) \right] ,
\end{equation}
where $R_{m}\equiv\hat{x}\cdot\vec{u}_{m}$, $L_{m}\equiv\hat{y}\cdot\vec
{u}_{m}$ ($\vec{u}_{m}=\vec{E}_{m}/\left| \vec{E}_{m}\right| $) and $%
\sum_{m=-l}^{l}\left[ \left| R_{m}\right| ^{2}+\left| L_{m}\right| ^{2}%
\right] $ $=1$. In this notation the \textit{basis vectors} $\left|
m,c\right) $ label the frequency and polarization, with $c=x,y$ and we can
associate $x$($y$) with the coin head (tail). We can also consider position
(or direction) cebits, that is, a field of the form

\begin{equation}
\vec{E}=\sum_{c=1,2}\sum_{m=-l}^{l}\vec{e}_{c}E_{m,c}e^{i\left[ \left(
\omega_{0}+m\bar{\omega}\right) t-k_{m}r_{c}\right] }+c.c.,
\end{equation}
\textit{\ }where $r_{c}=\hat{k}_{c}\cdot\vec{r}$ ($\hat{k}_{c}$ denotes the
direction of propagation of the $c^{th}$ field), and $\vec{e}_{c}$ denotes
its polarization. This can be represented by the state 
\begin{equation}
\left| \psi\right) =\sum_{m=-l}^{l}\left[ R_{m}\left| m,r_{1}\right)
+L_{m}\left| m,r_{2}\right) \right] ,
\end{equation}
where now $R_{m}\equiv E_{m,1}/\left( \left| E_{m,1}\right| ^{2}+\left|
E_{m,2}\right| ^{2}\right) $ and $L_{m}\equiv E_{m,2}/\left( \left|
E_{m,1}\right| ^{2}+\left| E_{m,2}\right| ^{2}\right) $. Generally we use
the index $c$ in $\left| m,c\right) $ to label the variable that plays the
role of the internal state of the coin; this is the polarization $c=x,y$ in
our first example, and the different paths, $c=r_{1},r_{2}$, in our second.
Of course more than two optical paths can be considered, and thus
coin--cebits with more that two components; this would be useful for the
implementation of multidimensional QWs \cite{Mackay}, but we shall not
consider these more complication dynamics here.

As stated, the walk is executed by the light frequency. Then there must be a
unitary operator that performs the conditional displacement operation 
\begin{equation}
\hat{V}\left| m,c_{i}\right) =\left| m\pm1,c_{i}\right) ,\;\;i=1,2.
\end{equation}
For polarization cebits, $\hat{V}$ can be physically implemented with an
electrooptic modulator (EOM) to which a linearly time dependent voltage is
applied: the $x$ ($y$) polarization component of the field frequency
component $\left( \omega+n\bar{\omega}\right) $ will see its frequency
increased (decreased) by an ammount $\bar{\omega}$. For a position cebit,
the EOMs must be adjusted so they are not polarization selective. Two such
EOMs are needed, one to increase the frequencies in one of the beams, and
the second to decrease the frequencies in the other beam.

After each displacement of the field frequency, a Hadamard transformation
acting on the cebit has to be implemented. The Hadamard transformation is
represented by the operator 
\begin{equation}
\hat{H}=\frac{1}{\sqrt{2}}\left( 
\begin{array}{cc}
1 & 1 \\ 
1 & -1
\end{array}
\right) ,
\end{equation}
whose action on the cebits reads 
\begin{align}
\hat{H}\left| m,c_{1}\right) & =\frac{1}{\sqrt{2}}\left[ \left|
m,c_{1}\right) +\left| m,c_{2}\right) \right] , \\
\hat{H}\left| m,c_{2}\right) & =\frac{1}{\sqrt{2}}\left[ \left|
m,c_{1}\right) -\left| m,c_{2}\right) \right] .
\end{align}
For polarization cebits, the Hadamard transformation is optically
implemented using a half--wave plate (HWP) with its fast axis forming an
angle $\pi/8$ with respect to the $\hat{x}$ axis. For position cebits, it is
implemented using a beam--splitter with two $\pi/2$ phase shifters \cite
{Spreeuw,Cerf}.

The QW is implemented by the repeated action on the cebit of the operator $%
\hat{H}\hat{V}$, that is, after $n$ iterations the state reads 
\begin{equation}
\left| \psi\right) _{n}=\left[ \hat{H}\hat{V}\right] ^{n}\left| \psi\right)
_{0},
\end{equation}
that can be written as 
\begin{equation}
\left| \psi\right) _{n}=\sum_{m=-n}^{+n}\left[ R_{m}\left| m,c_{1}\right)
+L_{m}\left| m,c_{2}\right) \right] ,
\end{equation}
with 
\begin{align}
R_{m,n} & =\frac{1}{\sqrt{2}}\left[ R_{m-1,n-1}+L_{m+1,n-1}\right] ,
\label{amp1} \\
L_{m,n} & =\frac{1}{\sqrt{2}}\left[ R_{m-1,n-1}-L_{m+1,n-1}\right] , 
\label{amp2}
\end{align}
where $R_{m,0}=L_{m,0}=0,\;$if$\;m\neq0$ and $R_{m,-1}=L_{m,-1}=0\;\forall m 
$. Finally, the intensity of each frequency component of the light field,
which is the optical analog of the probability of finding the walker at
position $m$ at iteration (time) $n$, is given by 
\begin{equation}
P_{m}\left( n\right) =\left| R_{m,n}\right| ^{2}+\left| L_{m,n}\right| ^{2}, 
\label{P}
\end{equation}
which is represented in Fig.1 $\footnote{%
We notice that the QW can also be defined as the repeated action of the
operator $\hat{V}\hat{H}$ as in \cite{Nayak}. In this case the probability
distribution is the same but the equations of evolution of the amplitudes
are different although can be easily related to the ones used here.}$.

In order to implement $n$ steps the best option is to introduce the
described elements in an optical cavity. In the next subsection we propose
several designs for optical cavities in which the QW on the line is
performed.

\subsection{Optical cavities for the QW on the line}

An optical cavity imposes the constraint that the optical frequencies must
fit within a set of eigenfrequencies defined \textit{e.g.} by the length of
the cavity. Thus, the time dependent electric field applied to the EOMs and
the cavity length must be adjusted in such a way that the frequency shift $%
\bar{\omega}$ matches the cavity free spectral range. We discuss this
important point below. It is also important to notice that the field
injected in the cavity from the outside (the initial state) can be injected
only once to effect the kind of initial conditions described above; more
complicated walks would be generated from multiple injections. In our
analysis we assume there are negligible cavity losses; in practice such
losses would limit the number of steps that could be implemented, unless an
amplifier were added to the cavity to compensate them. In Figs.3 and 4 we
illustrate in schematic form several possible designs for the implementation
of the QW on the line.

In Figs.3(a) and 3(b) a polarization cebit is considered. In the scheme of
Fig.3(a) we represent a ring cavity containing an EOM and a HWP, as
discussed in the previous section. In the scheme of Fig.3(b) the concept is
the same but utilizes a linear cavity. As a result, (i) the HWP has to be
replaced by a quarter wave--plate, QWP, and (ii) the frequency shift
introduced by the EOM must be half the size as compared to the ring cavity,
as light crosses both elements twice in every roundtrip.

Let us discuss in more detail the value of the frequency jump $\bar{\omega}$%
. We have commented above that the cavity imposses that $\bar{\omega}$ fits
the cavity free spectral range ($FSR$), but the frequency shift introduced
by the EOM cannot be better resolved than the inverse of the cavity
roundtrip time $\tau _{c}$, which is precisely $FSR$. Then the steps of the
QW would not be well resolved if their size equals $FSR$. Nevertheless there
are several ways of overcoming this difficulty. One way is to take the
frequency jump equal to several free spectral ranges, \textit{i.e.}, $\bar{%
\omega}=f$ $FSR$ with $f$ an integer larger than one. In this way the
uncertainty in the frequency displacement does not avoid the resolution of
the QW steps. Another possibility is that the frequency shift introduced by
the EOM be smaller than $FSR$. In this case it would be necessary several
roundtrips to perform a single frequency step of the QW and then it would be
necessary to control the action of the HWP (or the QWP) that performs the
Hadamard transformation as it should not act until the frequency step of the
QW is completed. For example, if the frequency step takes five cavity
roundtrips, the HWP should act only once every five rountrips. This can be
accomplished by substituting the HWP by a second EOM to which a constant
voltage of appropriate magnitude is applied, this voltage acting only every
five cavity roundtrips. There is still another possibility by using a single
pulse light field with a pulse duration shorter than $\tau _{c}$. In this
case $\bar{\omega}$ should be larger than the pulse spectral width but it
would be not necessary to match an integer of $FSR$. In any case, it is
clear that the sequence of frequency displacement and Hadamard
transformation can be performed within an optical cavity with enough
resolution.

\FRAME{fhFU}{7.686cm}{3.1148cm}{0pt}{\Qcb{{\protect\small Schemes for the
optical implementation of the QW on the line using polarization cebits. In
(a), a ring cavity, the electrooptic modulator (EOM) shifts up (down) the
frequency of the }$x${\protect\small \ (}$y${\protect\small ) polarization
component of the light field in }$\bar{\protect\omega}${\protect\small , and
a half-wave plate (HWP) with its axis forming an angle }$\protect\pi/8$%
{\protect\small \ with respect to the }$x${\protect\small -axis, performs
the Hadamard transformation. In (b), a linear cavity, the EOM modulator
shifts the field frequency }$\bar{\protect\omega}/2${\protect\small \
depending on the field frequency as in (a), and a quarter-wave plate (QWP)
performs the Hadamard transformation (notice that in the Fabry--Perot cavity
the light passes twice through each intracavity element every roundtrip). }}%
}{}{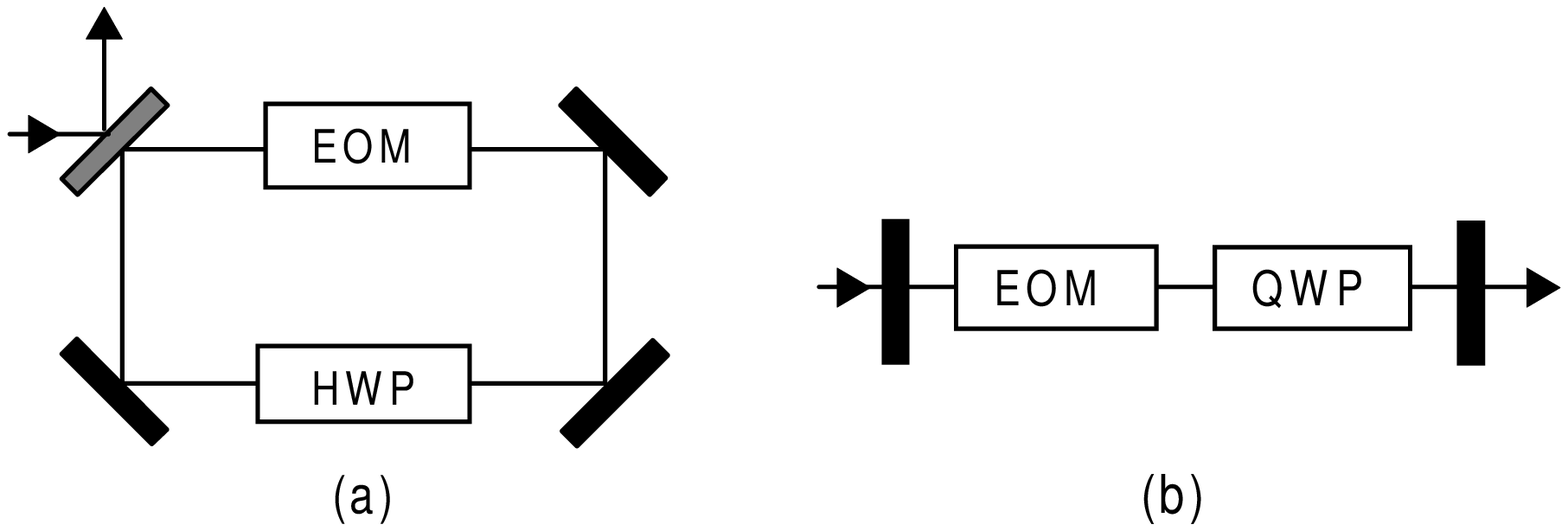}{\special{ language "Scientific Word"; type "GRAPHIC";
maintain-aspect-ratio TRUE; display "USEDEF"; valid_file "F"; width 7.686cm;
height 3.1148cm; depth 0pt; original-width 7.76in; original-height
11.0653in; cropleft "0.0696"; croptop "0.8520"; cropright "1"; cropbottom
"0.5920"; filename 'fig3.ps';file-properties "XNPEU";}}

In the ring cavity shown in Fig.4(a), a position cebit is used. Two
electrooptic modulators, EOM1 and EOM2, are needed here, each one of them
acting on each of the two sides of the cavity, which play the role of the
two coin states. The action of these EOMs must be polarization independent.
Alternately, the polarization of the light must be aligned with the axes of
the EOMs, and the set of axes of one EOM must be aligned with the set of
axes of the other. One of the EOMs increases the field frequency by $\bar{%
\omega}$, while the other decreases it by the same amount. The Hadamard
transformation is performed by the beamsplitter BS; we do not represent in
the figure the additional phase filters necessary for making a Hadamard, 
\cite{Spreeuw,Cerf}.

\FRAME{fhFU}{5.7881cm}{8.1539cm}{0pt}{\Qcb{{\protect\small Schemes for the
optical implementation of the QW on the line using position cebits. In (a),
two coupled unidirectional ring cavities, EOM1 and EOM2 increase and
decrease, respectively, the field frequency every roundtrip and the Hadamard
transformation is performed by the beamsplitter BS. The bidirectional ring
cavity in (b) is designed for sustaining a }$x${\protect\small \ (}$y$%
{\protect\small ) polarized field in the clockwise (counterclockwise)
direction. The EOM increases (decreases) the }$x${\protect\small \ (}$y$%
{\protect\small ) polarized field frequency and the set formed by the two
quarter--wave plates QWP1 (axis }$\protect\pi/4${\protect\small ), QWP2
(axis }$-\protect\pi/4${\protect\small ), and BS performs the Hadamard
transformation, see text.}}}{}{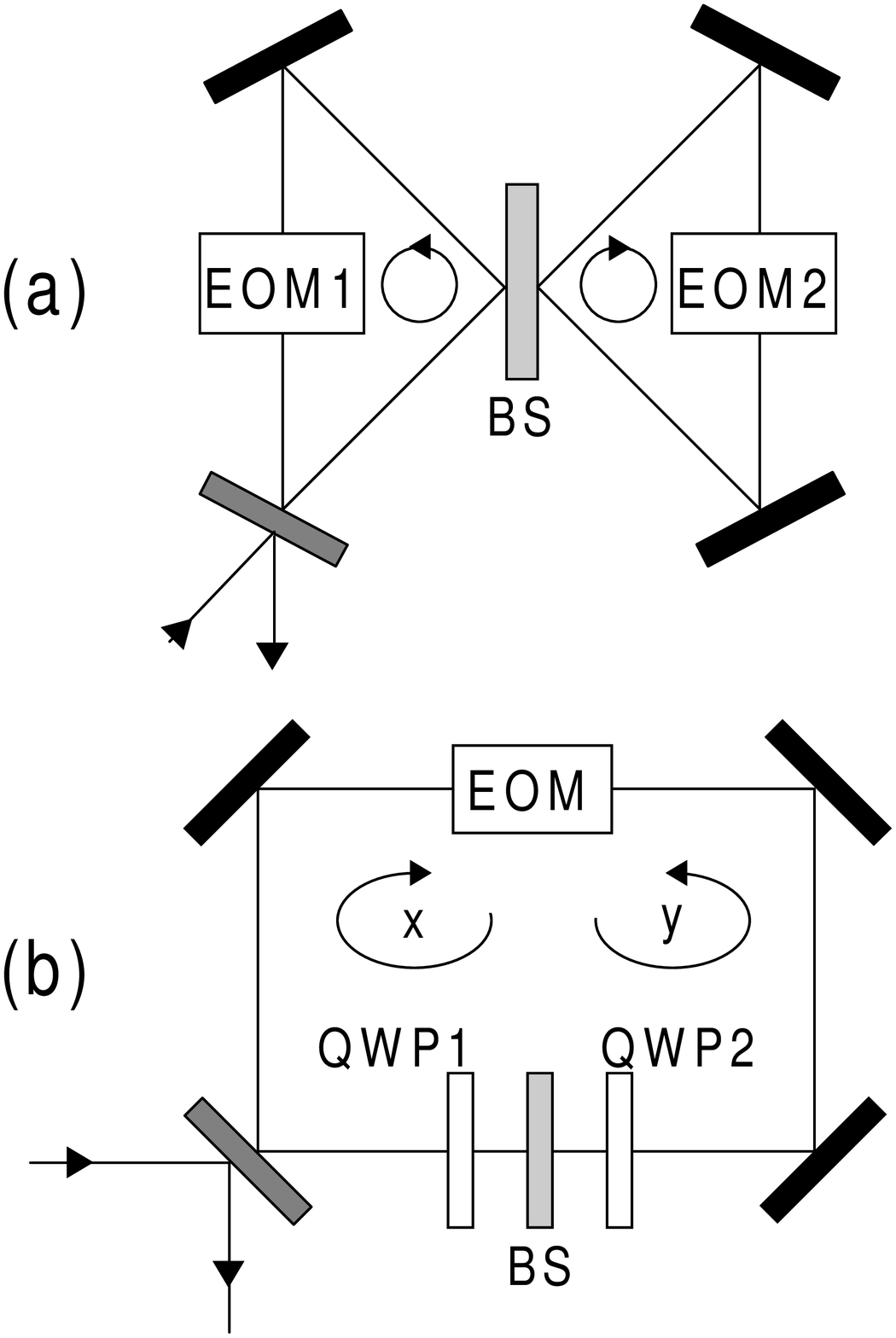}{\special{ language "Scientific
Word"; type "GRAPHIC"; maintain-aspect-ratio TRUE; display "USEDEF";
valid_file "F"; width 5.7881cm; height 8.1539cm; depth 0pt; original-width
7.76in; original-height 11.0653in; cropleft "-0.0102"; croptop "1";
cropright "1"; cropbottom "0"; filename 'fig4.ps';file-properties "XNPEU";}}

In Fig.4(b), again a ring cavity, a hybrid between a polarization and a
direction cebit is considered. In this bidirectional cavity the horizontal
and vertical polarization components of the field are forced to travel in
counterpropagating directions. As in the cases described in Fig.3, a single
EOM is needed that increases (decreases) the field frequency of the $x$ ($y$%
) polarization component in $\bar{\omega}$. The set of devices formed by the
two quarter--wave plates QWP1 (axis oriented at $\pi/4$) and QWP2 (axis
oriented at $-\pi/4$) and the BS plays a double function: (i) it guarantees
that the clockwise (counterclockwise) field remains $x$ ($y$) polarized, and
(ii) it performs the Hadamard transformation. The Hadamard transformation is
implemented by the BS; when done in the presence of the two QWP, this
guarantees that the transmitted field maintains its polarization, while the
reflected field changes its polarization from $x$ to $y$ and \textit{vice
versa}. Here is how it works: Note that when traversing the three elements
the polarization of a beam does not change, since the effect of QWP1 is
cancelled out by the effect of QWP2 , and \textit{vice versa}. For light
reflected by the BS, on the other hand, QWP1 (or QWP2) is crossed twice,
which is equivalent to the effect of a HWP with axis at $\pi/4$ ($-\pi/4$);
this changes $x$ polarization to $y$ polarization (or $y$ to $x$).

An interesting aspect of the two schemes shown in Fig.4 is that the output
corresponds to only one of the two cebit states and thus the spectrum of the
light does not correspond to $P_{m}\left( n\right) $, Eq.(\ref{P}), but to $%
P_{m}^{R}\left( n\right) =\left| R_{m,n}\right| ^{2}$ or $P_{m}^{L}\left(
n\right) =\left| L_{m,n}\right| ^{2}$. In order to obtain the complete QW,
one should allow the light to exit the cavity through two of the cavity
mirrors and then combine the two beams into a single beam. In Fig.5 we
represent $P_{m}\left( n\right) $ together with $P_{m}^{R}\left( n\right) $
and $P_{m}^{L}\left( n\right) $ for an initial condition different to that
of Fig.1, see caption.

\FRAME{fhFU}{6.4295cm}{9.138cm}{0pt}{\Qcb{{\protect\small Probability
distribution for }$n=200${\protect\small \ for the quantum walks. The
initial conditions chosen for calculating the QW were }$R_{0,0}=1$%
{\protect\small \ and }$L_{0,0}=0${{\protect\small , see Eqs.(\ref{amp1},\ref
{amp2}). Notice that }}$P_{m}${\protect\small \ is null for odd }$m$%
{\protect\small \ at odd }$n${\protect\small . We have represented only
nonzero values. }$P_{m}^{x},P_{m}^{y},${\protect\small \ and }$P_{m}$%
{\protect\small \ are represented in (a), (b), and (c), respectively.}}}{}{%
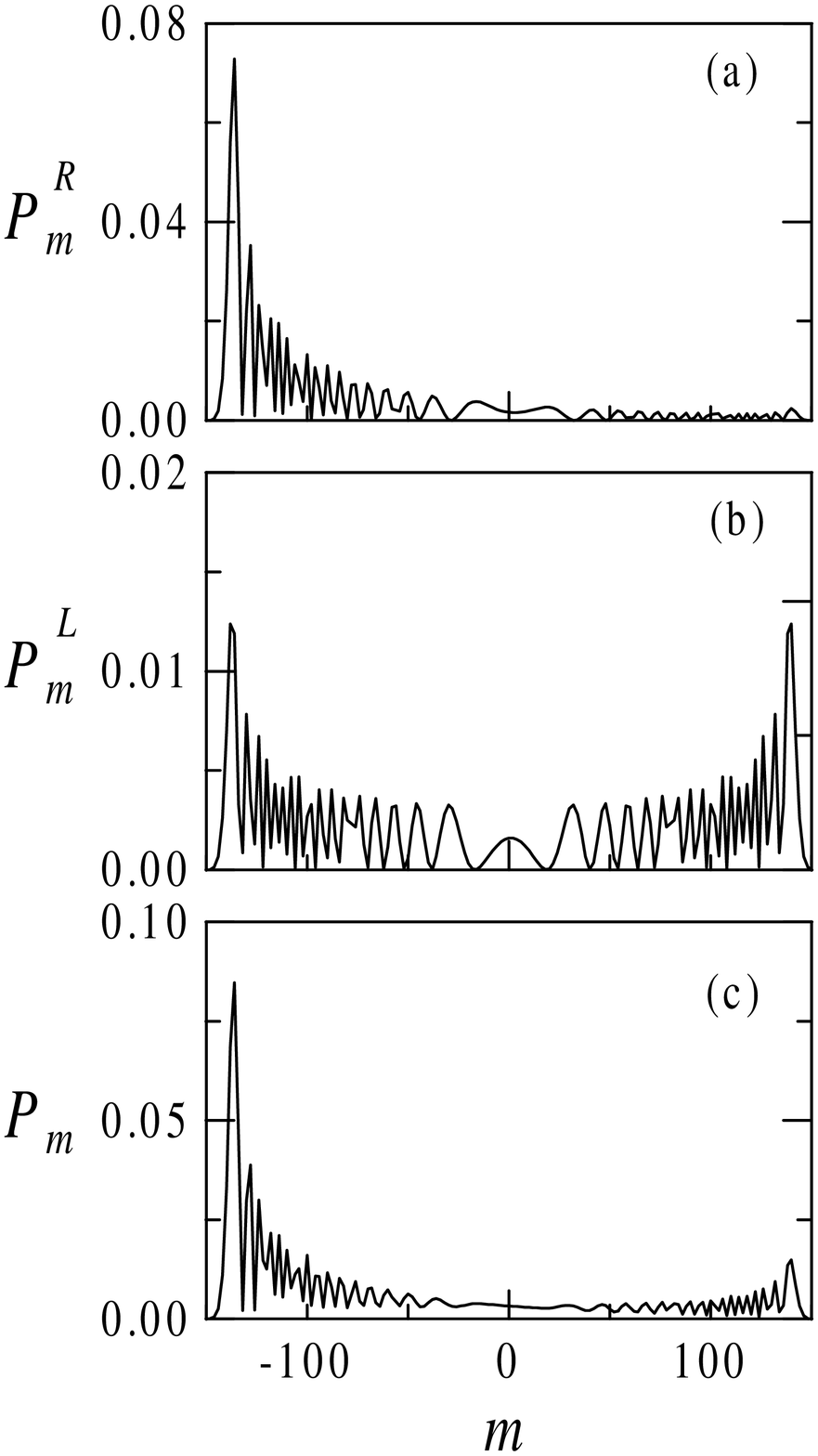}{\special{ language "Scientific Word"; type "GRAPHIC";
maintain-aspect-ratio TRUE; display "USEDEF"; valid_file "F"; width
6.4295cm; height 9.138cm; depth 0pt; original-width 7.76in; original-height
11.0653in; cropleft "0"; croptop "1.0014"; cropright "1.0014"; cropbottom
"0"; filename 'fig5.ps';file-properties "XNPEU";}}

\subsection{More general walks}

The Hadamard walk is only one of several possible ways of performing a QW.
What is actually needed is a unitary transformation that mixes the two
coin--cebit states after each system displacement but it is not necessary
that this unitary transformation be the Hadamard transformation. That is,
there is a family of unitary transformations that provide the same final
probability distribution, (see, \textit{e.g.}, \cite{Nayak,Konno,Bach}). In
particular, Konno \cite{Konno} has studied QWs for which the unitary
transformation applied to the qubit after each displacement reads 
\begin{equation}
U=\left( 
\begin{array}{cc}
a & b \\ 
c & d
\end{array}
\right) ,   \label{U}
\end{equation}
where $a,b,c$ and $d$ are complex, in general, and satisfy 
\begin{align}
\left| a\right| ^{2}+\left| b\right| ^{2} & =\left| c\right| ^{2}+\left|
d\right| ^{2}=1,  \notag \\
ac^{\ast}+bd^{\ast} & =0,  \notag \\
\left| \Delta\right| & =1,\Delta=ad-bc,  \notag \\
c & =-\Delta b^{\ast},d=\Delta a^{\ast}.
\end{align}
Konno \cite{Konno} demonstrated that, after a large number of iterations $n$%
, for QWs governed by such a $U$ the mean displacement of the walker, $%
\left\langle x\right\rangle $, and the mean quadratic displacement, $%
\left\langle x^{2}\right\rangle $, are given by 
\begin{align}
\left\langle x\right\rangle & =\left[ \left| \beta\right| ^{2}-\left|
\alpha\right| ^{2}+\frac{2\func{Re}\left( ab^{\ast}\alpha\beta
^{\ast}\right) }{\left| a\right| ^{2}}\right]  \notag \\
& \cdot\left( 1-\left| b\right| \right) n,  \label{xmean} \\
\left\langle x^{2}\right\rangle & =\left( 1-\left| b\right| \right) n^{2}, 
\label{x2mean}
\end{align}
where $\left| c\right) _{0}=col\left( \alpha,\beta\right) $ is the initial
state of the coin--qubit. The above results apply, obviously, to the
Hadamard walk by taking $a=b=c=-d=1/\sqrt{2}$.

For the implementations of the QW here considered, this means that it is not
essential that the unitary transformation after each frequency displacement
be implemented by a HWP, in the case of polarization cebits, nor by a 50/50
beamsplitter for position cebits, as they can be substituted by any unitary
transformation of the type $U$ on the polarization or position. Now we have
the elements required for discussing the experiemnt of Ref.\cite{Bouwmeester}%
.

\subsection{The experiment of Bouwmeester et al. \protect\cite{Bouwmeester}}

Bouwmeester \cite{Bouwmeester} proposed an optical implementation of the
Galton board (the quincunx), and studied it both theoretically and
experimentally. What was actually implemented is a grid of Landau-Zener
crossings through which a light beam propagates, and the focus was the study
of the appearance of recurrences in the spectrum of the light field. We
refer the interested reader to Ref.\cite{Bouwmeester} for details.

A simplified version of the experimental device used in \cite{Bouwmeester}
is that represented in Fig.3(b), but replacing the QWP by a second EOM with
its axis rotated $\pi/4$ with respect to the axis of the first EOM. The
second EOM introduces a dephasing $\delta/2$ between the two polarization
components, where $\delta$ can be varied at will. Then, after a cavity
roundtrip, the matrix representing the action of the second EOM action on
the $x$ and $y$ polarization components of the field reads 
\begin{equation}
U_{\delta}=\left( 
\begin{array}{cc}
\cos\delta & -i\sin\delta \\ 
-i\sin\delta & \cos\delta
\end{array}
\right) .   \label{Udelta}
\end{equation}
It is easy to see that $U_{\delta}$ can be seen as a particular case of $U$ (%
\ref{U}) for $a=d=\cos\delta$ and $b=c=-i\sin\delta$. From Eqs.(\ref{xmean})
and (\ref{x2mean}) we find 
\begin{align}
\left\langle x\right\rangle & =\left[ \left| \beta\right| ^{2}-\left|
\alpha\right| ^{2}+2\func{Im}\left( \alpha\beta^{*}\right) \tan\delta\right]
\notag \\
& \cdot\left( 1-\sin\delta\right) n, \\
\left\langle x^{2}\right\rangle & =\left( 1-\sin\delta\right) n^{2}.
\end{align}
Notice that for $\alpha=\beta$ both real quantities -- that is, for an
initial linearly polarized state -- $\left\langle x\right\rangle =0$ for all 
$\delta$ and the variance of the spectrum is $\sigma=\sqrt{1-\sin\delta}$.

From this it is clear that the experimental device used by Bouwmeester \cite
{Bouwmeester} can implement QWs. In fact, in Fig.6 of their work \cite
{Bouwmeester}, which corresponds to $\delta=\pi/5$, a power spectrum very
similar to those expected for the optical implementation of the QW can be
clearly appreciated. The authors \cite{Bouwmeester} considered Fig.6 in
their manuscript as a demonstration of the coherence quality of their
system, and did not make an explicit connection of their work with QWs.

Nevertheless there is subtle but essential difference between the
experiement of \cite{Bouwmeester} and an actual implementation of the QW. In 
\cite{Bouwmeester} the frequency displacement $\bar{\omega}$ is a fraction
of the $FSR$ but the unitary transformation (18) is applied at every
roundtrip (see our discussion above in Subsection 3.2). Then, the experiment
in \cite{Bouwmeester} cannot be regarded as an implementation of the QW
although is something certainly very similar. That is, the optical Galton
board is different from the QW. Nevertheless we can consider the experiment
of Bouwmeester \textit{et al.} \cite{Bouwmeester} as a proof of principle
that the QW can be implemented in an optical cavity.

\subsection{Quantum walk on the circle}

The QW on the circle, again with the frequency playing the role of position,
can be also implemented optically. If $2M-1$ is the number of possible
discrete values for the frequency, it is necessary to devise a way of
shifting the frequency $\omega+\left( M+1\right) \bar{\omega}$ to $\omega -M%
\bar{\omega}$ and the frequency $\omega-\left( M+1\right) \bar{\omega}$ to $%
\omega+M\bar{\omega}$. This can be done by substituting the EOMs in the
schemes of Figs.3 and 4 by the device shown in Fig.6, which we denote as $%
\overline{EOM}$. It consists of two electrooptic modulators and two
specially designed mirrors. Mirrors M1 and M2 only reflect the frequencies $%
\omega +M\bar{\omega}$ and $\omega-M\bar{\omega}$, respectively, and are
transparent to the rest of the frequencies in the spectrum.

\FRAME{fhFU}{6.9281cm}{3.7101cm}{0pt}{\Qcb{{\protect\small Scheme of the }$%
\overline{EOM}${\protect\small \ device that substitutes the electrooptic
modulators in Fig.3(a) in order to perform the QW on the circle. M1 and M2
are mirrors that reflect a single frequency (}$M\bar{\protect\omega}$%
{\protect\small \ and }$-M\bar {\protect\omega}${\protect\small ,
respectively) and EOMa and EOMb are electrooptic modulators. See text for a
detailed description of the device operation. }}}{}{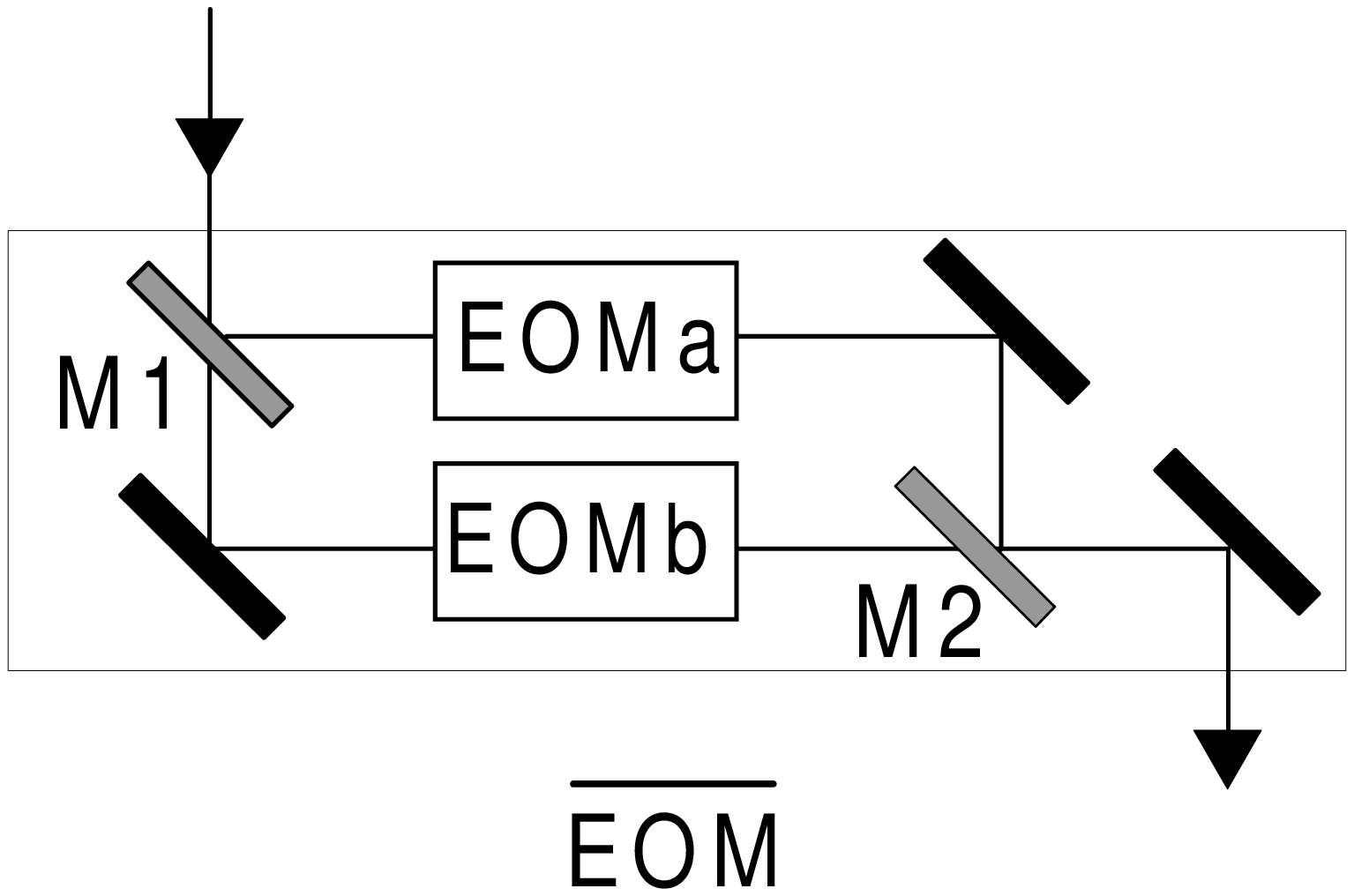}{\special{
language "Scientific Word"; type "GRAPHIC"; maintain-aspect-ratio TRUE;
display "USEDEF"; valid_file "F"; width 6.9281cm; height 3.7101cm; depth
0pt; original-width 19.7103cm; original-height 28.1058cm; cropleft "0";
croptop "0.7525"; cropright "0.9985"; cropbottom "0.3809"; filename
'fig6.ps';file-properties "XNPEU";}}To see how the circle walk results, we
consider as an example the scheme in Fig. 4(a). First we look at the left
side of the cavity, in which frequencies were increased at EOM1 in the QW on
the line. With $\overline{EOM}$ of Fig. 6 replacing EOM1 in Fig.4(a),
consider any step at which the field entering $\overline{EOM}$ contains all
the allowed frequencies from $\omega -M\bar{\omega}$ to $\omega+M\bar{\omega}
$. The role of M1 is to separate the frequency $\omega+M\bar{\omega}$, which
is directed to EOMa, from the rest of frequencies, which are directed to
EOMb. The modulators are configured such that EOMa decreases the entering
frequency by $2M\bar{\omega},$ while EOMb increases the frequencies entering
it in $\bar{\omega}$. After traversing the electrooptic modulators, the
frequencies enter M2. The frequencies $\left\{ \omega-\left( M-1\right) \bar{%
\omega},\ldots,\omega+M\bar{\omega}\right\} $ that come from EOMb traverse
M2, while the frequency $\omega-M\bar{\omega}$, that comes from EOMa is
reflected by M2. Then, the set of frequencies exiting $\overline{EOM} $ is
the same set that entered, but all of them have been shifted appropriately.

In the right side of the cavity in Fig.4(a), where frequencies were
decreased in the QW on the line, EOM2 is substituted by another $\overline{%
EOM}$ that works in a slightly different way to the one in the left side of
the cavity: first, the positions of M2 and M1 are interchanged (or, what is
the same, light enters from below), and second, EOMa and EOMb in $\overline{%
EOM}$ are modified in the sense that EOMa now \textit{increases} the
incoming frequency in $2M\bar{\omega}$ whilst EOMb \textit{decreases} in $%
\bar{\omega}$ the rest of frequencies. In this way, in the right side of the
cavity all frequencies entering $\overline{EOM}$ are decreased except $%
\omega-M\bar{\omega}$ that is converted into $\omega+M\bar{\omega}$ as it
must be. Of course it is essential that the optical lengths of the two paths
in $\overline{EOM}$ be symmetric in order not to introduce any spurious
dephasing during the separation of frequencies.

We finally comment that the device $\overline{EOM}$ can be also used in the
cavities of Figs.3(a) and 4(b) by suitably adapting its operation. In
particular, in the scheme of Fig.4(b) a single $\overline{EOM}$ is
sufficient for performing the QW on the circle: as the clockwise and
counterclockwise fields have orthogonal polarizations, the electrooptic
modulators in $\overline{EOM}$ can increase (decrease) the frequencies
oppositely for the two polarization components.

\section{Discussion}

We have shown in detail how quantum walks, both on the line and on the
circle, can be implemented in optical cavities. We have also discussed the
close relation existing between the optical Galton board of Bouwmeester 
\textit{et al.} \cite{Bouwmeester} and the QW on the line. In fact we have
shown that the experiment of \cite{Bouwmeester} can be regarded as a proof
of principle that our proposals for the implementation of the QW can be
carried out in optical cavities. All of this leads to the obvious
conclusion, as noted earlier by us \cite{Knight}, that the coined QW does
not require quantization for its implementation and thus, in this sense, it
is not quantum. Nevertheless this point requires further discussion.

First of all we should carefully distinguish between an actual
implementation and a simulation of the QW. A simulation would not capture
all the characteristics of the QW whilst an implementation would do. Then,
in order to make an assesment as strong as ''the QW is not quantum but
classical'' we must be completely sure that what we are proposing is an
implementation and not a simulation as in this last case some quantum
features would be lost an this would manifest in some measurable
characteristics. In order to make this distinction clearer, let us consider
the optical \textit{implementations} of the Grover's search algorithm \cite
{Kwiat,Bhattacharya}: it is rather clear that these cannot be regarded as
actual implementations but only as simulations as the lack of quantum
entanglement manifests in the exponential increase of classical resources
with the increase of the problem size. In other words, with the only use of
classical resources, quantum entanglement can be simulated but not
implemented and this manifests in the scaling of the needed resources. Then,
our proposal is an implementation or a simulation?

In order to answer this question we should first clarify the role played by
entanglement in the QW, something that is beyond the scope of this article
and that we plan to investigate elsewhere, but even without this knowledge
we can still clarify to some extent this point. It is obvious that the
classical resources needed in the schemes here proposed for the
implementation of the QW on the line (or on the circle) do not grow with the
number of steps in the QW except, perhaps, in the detection stage but here
the resources needed would grow linearly with the number of steps, as would
grow in a quantum implementation. Then it seems that entanglement does not
play any role in the QW and we can consider our proposals as actual
implementations. But this first conclussion could be misleading as the
increase in the problem size does not lie in the number of steps in the walk
but in the \textit{dimension} of the QW. We have noted above that our
schemes could be generalized for implementing multidimensional QWs \cite
{Mackay}:\ for that, coins with more than two states are needed or, in other
words, several coins are needed. Polarization and direction cebits could be
used simultaneously for implementing these multidimensional QWs. It would be
here where the problem of the scaling of the classical resources would
appear: a $d$ dimensional QW requires $d$ coin-qubits but their classical
simulation requires $2^{d}$ light beams \cite{Spreeuw}. Then, it is in these
higher dimensional walks that quantum entanglement would manifest. In
conclussion:\ one--dimensional (even two--dimensional) QWs can be
classically implemented but higher dimensional QWs can only be simulated. Of
course this is a very preliminary conclussion as in some cases
multidimensional QWs could be reduced to equivalent unidimensional QWs \cite
{KempeCP}.

Let us finally note that one of the interesting phenomena that could be
studied by using any of the designs here proposed is the effect of
decoherence on the QW. It has been shown by Kendon and Tregenna \cite{Kendon}
that decoherence can be beneficial for QWs in the sense that it can lead,
under suitable conditions, to more homogeneous probability distributions.
Decoherence can act on the walker and/or on the coin, and it could be
simulated in the schemes devised here by introducing, \textit{e.g.}, random
changes in the polarization or direction of the light beams.

\section{Acknowledgements}

This work has been supported in part by the UK Engineering and Physical
Sciences Research Council and the European Union. ER acknowledges a grant
from the Ministerio de Educaci\'{o}n, Cultura y Deportes of the Spanish
Goverment (Grant PR20002-0244) and partial finantial support from the
spanish Ministerio de Ciencia y Tecnolog\'{i}a and European FEDER funds
through project BFM2002-04369-C04-01. JES acknowledges financial support
from the Natural Sciences and Engineering Research Council of Canada. We
gratefully acknowledge Dr. Javier Garc\'{i}a (Universitat de Val\`{e}ncia,
Spain) for his careful reading of the manuscript.


\begin{thebibliography}{99}
\bibitem{Spreeuw}  R.J.C. Spreeuw, Phys. Rev. A \textbf{63}, 062302 (2001)

\bibitem{Cerf}  N.J. Cerf, C. Adami, and P.G. Kwiat, Phys. Rev. A \textbf{57}%
, R1477 (1998)

\bibitem{Kwiat}  P.G. Kwiat, J.R. Mitchell, P.D.D. Schwindt, and A.G. White,
J. Mod. Opt. \textbf{47}, 157 (2000)

\bibitem{Bhattacharya}  N. Bhattacharya, H.B. van Linden van den Heuvell,
and R.J.C. Spreeuw, Phys. Rev. Lett. \textbf{88}, 137901 (2002)

\bibitem{Aharonov}  Y. Aharonov, L. Davidovich, and N. Zagury, Phys. rev. A 
\textbf{48}, 1687 (1993)

\bibitem{Meyer}  D. Meyer, J. Stat. Phys. \textbf{85}, 551 (1996)

\bibitem{Farhi}  E. Farhi and S. Gutman, Phys. Rev. A \textbf{58}, 915 (1998)

\bibitem{Nayak}  A. Nayak, and A. Vishwanath, e-print quant-ph/0010117

\bibitem{DAharonov}  D. Aharonov, A. Ambainis, J. Kempe, and U. Vazirani,
Proceedings of the 30th Anual ACM Symposium on Theory of Computation (ACM
Press, New York, 2001) 50; e-print quant-ph/0012090

\bibitem{Kempe}  J. Kempe, e-print quant-ph/0205083

\bibitem{Konno}  N. Konno, Quantum Information Processing \textbf{1}, 345
(2002); e-print quant-ph/0206103

\bibitem{Bach}  E. Bach, S. Coppersmith, M.P. Goldschen, R. Joynt, and J.
Watrous, e-print quant-ph/0207008

\bibitem{Mackay}  T.D. Mackay, S.D. Barltett, L.T. Stephenson, and B.C.
Sanders, J. Phys. A: Math. Gen. \textbf{35}, 2745 (2002)

\bibitem{Shenvi}  N. Shenvi, J. Kempe, and K.B. Whaley, Phys. Rev. A \textbf{%
67}, 052307 (2003)

\bibitem{Childs}  A.M. Childs, R. Cleve, E. Deotto, E. Farhi, S. Gutman, and
D. A. Spielman, quant-ph/0209131

\bibitem{Kendon}  V. Kendon and B. Tregenna, Phys. Rev. A \textbf{67},
042315 (2003)

\bibitem{Bednarska}  M. Bednarska, A. Grudka, P. Kurzynski, T. Luczak, and
A. Wojcik, e-print quant-ph/0304113

\bibitem{Hillery}  M. Hillery, J. Bergou, and E. Feldman, e-print
quant-ph/0302161

\bibitem{Carteret}  H.A. Carteret, M.E.H. Ismail, and B. Richmond, e-print
quant-ph/0303105

\bibitem{Tregenna}  B. Tregenna, W. Flanagan, R. Maile, and V. Kendon,
e-print quant-ph/0304204

\bibitem{KempeCP}  J. Kempe, Contemp. Phys. \textbf{44}, 307 (2003)

\bibitem{Bartlett}  We thank Steven Bartlett, of Macquarie University, for
pointing this out to us. See, for example, C.H. Papadimitriou, \textit{%
Computational Complexity} (Addison Wesley, 1994)

\bibitem{Travaglione}  B.C. Travaglione and G.J. Milburn, Phys. Rev. A 
\textbf{65}, 032310 (2002)

\bibitem{Dür}  W. D\"{u}r, R. Raussendorf, V.M. Kendon, and H.-J. Briegel,
Phys. Rev. A \textbf{66}, 052319 (2002)

\bibitem{Sanders}  B.C. Sanders, S.D. Bartlett, B. Tregenna, and P.L.
Knight, Phys. Rev. A \textbf{67}, 042305 (2003)

\bibitem{Zhao}  Z. Zhao, J. Lu, H. Li, T. Yang, Z.-B Chen, and J.-W. Pan,
e-print quant-ph/0212149

\bibitem{Jeong}  H. Jeong, M. Paternostro, and M.S. Kim, e-print
quant-ph/0305008

\bibitem{Knight}  P.L. Knight, E. Rold\'{a}n, and J.E. Sipe, Phys. Rev. A 
\textbf{68}, 020301 (2003)

\bibitem{Bouwmeester}  D. Bouwmeester, I. Marzoli, G.P. Karman, W. Schleich,
and J.P. Woerdman, Phys. Rev. A \textbf{61}, 013410 (2000)
\end{thebibliography}
\end{document}